\pdfoutput=1
\documentclass[aps,prl,reprint,superscriptaddress]{revtex4-2}
\usepackage{amsmath} 
\usepackage{amssymb} 
\usepackage{physics} 
\usepackage{graphicx} 
\usepackage{hyperref} 

\makeatletter
\newcommand{\colorcaption}[2][]{%
  \begingroup%
  \renewcommand{\@caption@fignum@sep}{ (Color online). }%
  \caption[#1]{#2}%
  \endgroup%
}
\makeatother

\begin{document}

\title{Anisotropy-Exchange Resonance as a Mechanism for Entangled State Switching}

\author{Eric D. Switzer}
\affiliation{Department of Physics, University of Central Florida, Orlando, Florida 32816, USA}
\author{Xiao-Guang Zhang}
\affiliation{Department of Physics, Center for Molecular Magnetic Quantum Materials and Quantum Theory Project, University of Florida, Gainesville, Florida 32611, USA}
\author{Talat S. Rahman}
\email[Corresponding author email: ]{talat.rahman@ucf.edu}
\affiliation{Department of Physics, University of Central Florida, Orlando, Florida 32816, USA}

\date{\today}

\begin{abstract}
We explore the three-particle spin model of an $S_{1}=\frac{1}{2}$ particle (e.g. a stationary electron) interacting with two spin-coupled $S_{\text{2,3}}$ particles with exchange coupling and magnetic anisotropy. 
We find that in the case of $S_{2,3}=1$ particles, the coupled particle entanglement states can be prepared, controlled, and read by the $S_{1}$ particle. 
We also find that for particular resonance conditions of the magnetic anisotropy strength $D$ and exchange coupling strength $J$, the entanglement state switching behavior is maximized and is robust against a range of anisotropic application of the exchange coupling.
\end{abstract}

\maketitle

    Spin state entanglement plays a key role in many systems, including those considered within quantum information science (QIS). 
    For example, spin qubits, the coherent superposition of spin states within quantum objects, make use of entangled spin states for quantum gate operations \cite{nielsenbook}. 
    Spin qubits have been explored theoretically and experimentally, notably in the application of confined electrons in quantum dots fabricated in semi-conductors \cite{loss98,petta05,hanson07,noiri16,noiri18,nakajima19,yang2020,leon2020} and the search for robust QIS-applicable magnetic molecule systems \cite{leuenberger01, vandersypen01,vincent12,ganzhorn13,urdampilleta13,thiele14,pedersen16,najafi19}. 
    Molecular magnets in particular possess an onsite magnetic anisotropy which gives rise to their unique magnetic properties. Molecular magnets are viable candidates for spin state switching for QIS purposes because of their long coherence times, the ability to tunnel between spin states resulting from their magnetic anisotropy, and tailorable ligands \cite{bogani08}. 
    For example, the single molecule magnet $\text{TbPc}_{2}$ possesses a nuclear spin that is electrically controllable and has long coherence times \cite{thiele14,najafi19}. In both of these QIS approaches, the Kondo effect has been found \cite{cronenwett98,leuenberger06,gonzalez08}, and thus the Kondo or Anderson impurity model \cite{kondo64,anderson61} is applicable to predict some of the features of these systems. 
    
    With these considerations, a QIS system that contains onsite magnetic anisotropy is expected to have a complex interaction between the system's anisotropy and effective exchange coupling.
    While some studies have examined the interplay of exchange coupling and onsite magnetic anisotropy for two particles \cite{zitko08}, the three-particle case is a qualitatively different system that has not been fully explored.
    Some experimental and theoretical studies have realized multiple-quantum dot scenarios \cite{mehl15,noiri16,noiri18,nakajima19}, or studied the two magnetic impurity entanglement state dependency of contact exchange interactions with incident electrons \cite{costa06,ciccarello06,ciccarello09}. 
    As described in the effectively three spin particle setup in Ref.~\cite{hiraoka17}, the strong-coupling Kondo exchange regime and the weak-coupling spin-orbit interaction regime compete with each other, resulting in a non-trivial interaction. In all of these studies, the exchange coupling mechanism plays a significant role in controlling the system of interest. Unintended variations in this exchange can cause undesirable effects, and thus a system must be correspondingly robust against them. 

    In this work, we explore a general spin model with exchange and magnetic anisotropy that encompasses these scenarios and investigates the robustness of the spin system by extending the two-particle work of Ref.~\cite{zitko08} to the three-particle paradigm. 
    We consider two magnetic sites of either $S_{2,3}=\frac{1}{2}$ or $S_{2,3}=1$ in which a Kondo-like exchange interaction is applied either isotropically or anisotropically between them and the $S_{1}=\frac{1}{2}$ particle. 
    Because we do not treat all degrees of freedom (e.g. the electronic degrees of freedom) and instead focus solely on the model's spin degrees of freedom, we consider a model that has a physical analog with the recently realized experiments of three quantum dots \cite{noiri16,noiri18,nakajima19}. 
    As we will show, we find that for the $\mathbb{C}^{18}$ state space model corresponding to $S_{2,3}=1$, the exchange and anisotropy interactions lead to a set of conditions on the exchange and magnetic anisotropy strengths that correspond with perfect non-entangled to entangled state switching in four smaller SO(2) representation sub-groups.
    We find that at these special resonance conditions, which we designate as ``DJ resonances,'' measurement of the coupled particle entanglement states is possible by measurement of the $S_{1}=\frac{1}{2}$ particle's spin.
    We also show the conditions in which these DJ resonances allow for complete control of appropriately-chosen Bloch vectors within a subspace of the coupled particles' total spin space, which is not found for the $S_{2,3}=\frac{1}{2}$ model. 
    We demonstrate conditions for full control of this Bloch vector, and that state coherence is robust against anisotropic application of the exchange coupling.

\begin{figure}
    \includegraphics[width=150pt]{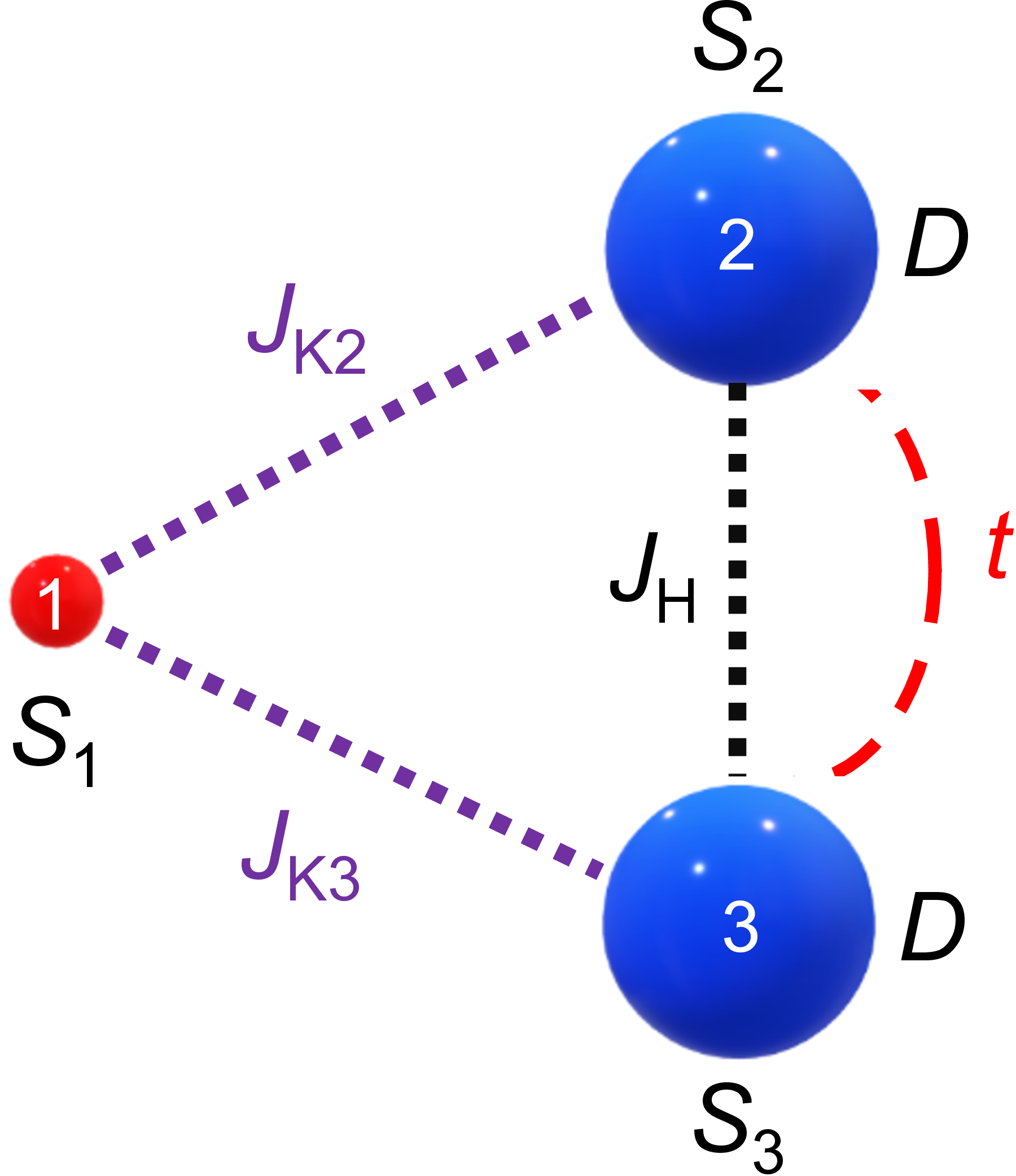}
    \colorcaption{\label{fig:model-schematic}
    Schematic of the spin model considered in this work. 
    Particle 2 and 3 are coupled by an exchange interaction $J_{\text{H}}$. 
    Particle 1 is also coupled to particle 2 and 3 by an exchange interaction, $J_{\text{K2}}$ and $J_{\text{K3}}$, respectively. 
    Particle 1 is allowed to tunnel between particle 2 and particle 3 with hopping strength $t$.}
    \end{figure}

    A representative schematic of the resulting spin model is shown in Fig.~\ref{fig:model-schematic}. 
    The spin Hamiltonian ($\hbar$ = 1) is then,
    \begin{align}
    \label{eqn:hamiltonian-total}
        \mathcal{H} &= \mathcal{H}_{\text{H}}
                      +\mathcal{H}_{\text{K}}
                      +\mathcal{H}_{\text{A}}
                      +\mathcal{H}_{\text{T}},
    \end{align}
    where each term in the Hamiltonian is explained as follows. 
    The Heisenberg-like interaction of the two $S_{2,3}$ particles is represented by 
    \begin{align}
    \label{eqn:hamiltonian-heisenberg-interaction}
        \mathcal{H}_{\text{H}} = 
        J_{\text{H}}\hat{\vb{S}}_{2}\cdot\hat{\vb{S}}_{3},
    \end{align}
    where $\hat{\vb{S}}_{i}=(\hat{S}^{x}_{i},\hat{S}^{y}_{i},\hat{S}^{z}_{i})$ is the spin operator for the $i$th particle, and $J_{\text{H}}$ is the strength of the exchange interaction. 
    The Kondo-like interaction of the $S=\frac{1}{2}$ particle with the $S_{2,3}$ particles is represented by,
    \begin{align}
    \label{eqn:hamiltonian-kondo-interaction}
        \mathcal{H}_{\text{K}}=\frac{1}{2}\sum_{\mu,\mu^{\prime},i=2,3}J_{\text{Ki}}\hat{\vb{S}}_{i}\cdot\hat{\vb{d}}^{\dagger}_{\mu,i}\hat{\bm{\sigma}}_{\mu,\mu^{\prime}}\hat{\vb{d}}_{\mu^{\prime},i},
    \end{align}
    where $\mu$/$\mu^{\prime}$ is a spin index for particle 1, $\vb{\sigma}_{\mu,\mu^{\prime}}$ is the corresponding $\mu,\mu^{\prime}$ matrix element of the $s=\frac{1}{2}$ Pauli matrix, and $\hat{\vb{d}}^{\dagger}_{\mu,k}$/$\hat{\vb{d}}_{\mu,i}$ represents (in second quantization language) the creation/annihilation operator of a state in which particle 1 is bound to particle $i$. 
    In our general treatment, we allow $J_{\text{K2}}$ and $J_{\text{K3}}$ to take all values, i.e. we consider both ferromagnetic and anti-ferromagnetic possibilities.
    Additionally, we consider situations in which $S_{2,3} = 1$ particles possess an anisotropic response to applied magnetic fields,
    \begin{align}
    \label{eqn:hamiltonian-anisotropy}
        \mathcal{H}_{A}=D\left(\hat{S}^{z}_{2}\hat{S}^{z}_{2}+\hat{S}^{z}_{3}\hat{S}^{z}_{3}\right), 
    \end{align}
    where $D$ is a uniaxial anisotropy strength. 
    Again, our general treatment permits D to span all values, which allows one to consider to both ``easy-axis'' and ``hard-axis'' anisotropies. 
    Last, the tunneling Hamiltonian is described by,
    \begin{align}
    \label{eqn:hamiltonian-tunneling}
        \mathcal{H}_{\text{T}} = 
        \sum_{\mu}\left\{t\hat{\vb{d}}^{\dagger}_{\mu,2}\hat{\vb{d}}_{\mu,3}+h.c.\right\},
    \end{align}
    where $\mu$ is the spin index for particle 1.
    The dynamics of the system, which may involve entangled particle scenarios,
    is described by the density operator $\rho$ in the Schr\"{o}dinger picture,
    \begin{align}
    \label{eqn:liouville-von-neumann}
        i\pdv{\rho}{t}=\left[\mathcal{H},\rho\right],
    \end{align}
    where the brackets denote the commutator. 

    There are various basis sets that uncover different aspects of the dynamics of the three-spin system. 
    One of these is the spin product states that are aligned to the action of the $\hat{S}^{z}=\hat{S}^{z}_{1}\otimes\hat{S}^{z}_{2}\otimes\hat{S}^{z}_{3}$ operator, i.e. the Hamiltonian and density operator are represented in the product basis $\ket{s,m_{s}}=\ket{s_{1},m_{s_{1}}}\otimes(\ket{s_{2},m_{s_{2}}}\otimes\ket{s_{3},m_{s_{3}}})$, where $\vb{S}=\vb{S}_{1}+\vb{S}_{2}+\vb{S}_{3}$. 
    Because we also need to examine the possible entangled states of the $S_{2,3}$ particles in anticipation of correlating states within a qubit representation, we designate a ``device'' basis which is given by $\ket{s_{1},m_{1}}\otimes\ket{s_{23},m_{23}}$. 
    In this representation, $\ket{s_{23},m_{23}} = \ket{s_{2},m_{2}}\otimes\ket{s_{3},m_{3}}$ is designated the ``coupled particle'' basis.
    
    We first consider the $S_{2,3}=1$ model and the impact of each term within the total Hamiltonian of Eq.~(\ref{eqn:hamiltonian-total}) on the states of the system. 
    When the Kondo-like exchange Hamiltonian is applied anisotropically (i.e. $J_{\text{K2}} \ne J_{\text{K3}}$), the Hamiltonian connects states between different $s_{23}$ subspaces in the device representation and can no longer be block-diagonalized by the $s_{23}$ subspaces. 
    Instead, the effective exchange Hamiltonian can be block diagonalized by $m$ values, where $\mathcal{H}_{m}$ is the block Hamiltonian corresponding to $m$, and $\mathcal{H}_{\pm 5/2}$ are diagonal. In the
    $\mathcal{H}_{\pm 3/2}$ subspaces, the $\ket{m_{1}}\ket{s_{23},m_{23}}=\ket{\pm\tfrac{1}{2}}\ket{2,\pm1}$, $\ket{\pm\tfrac{1}{2}}\ket{1,\pm1}$, and $\ket{\mp\tfrac{1}{2}}\ket{2,\pm2}$ states participate, forming three-dimensional subspaces. 
    Similarly, the $\mathcal{H}_{\pm 1/2}$ subspaces contains the interactions of the $\ket{\pm\tfrac{1}{2}}\ket{2,0}$, $\ket{\pm\tfrac{1}{2}}\ket{1,0}$, $\ket{\pm\tfrac{1}{2}}\ket{0,0}$, $\ket{\mp\tfrac{1}{2}}\ket{2,\pm1}$, and $\ket{\mp\tfrac{1}{2}}\ket{1,\pm1}$ states, making the subspaces five dimensional.
    These forms of the effective exchange Hamiltonian will play a pivotal role in transitions between states with the same $m$ value. 
    
    In the $S_{2,3}=1$ model, resonant transitions between states are found in several of the $m$ subspaces. 
    By inspecting the $m=\frac{3}{2}$ subspace, which corresponds with the dynamics of the $\ket{m_{1}}\ket{s_{23},m_{23}}=\left\{\ket{\uparrow}\ket{2,1},\ket{\uparrow}\ket{1,1},\ket{\downarrow}\ket{2,2}\right\}$ states, the block Hamiltonian takes the form (a common $t+t^{*}+J_{\text{H}}+D+\frac{1}{4}\Sigma_{\text{K}}$ is removed from the diagonal from this point on),
    \begin{align}
    \label{eqn:hamiltonian-subblock-3d2-devicebasis}
        \mathcal{H}_{3/2}&=\frac{1}{4}
        \begin{pmatrix}
            0 & \Delta_{\text{K}} & 2\Sigma_{\text{K}}\\
            \Delta_{\text{K}} & -8J_{H} & -2\Delta_{\text{K}}\\
            2\Sigma_{\text{K}} & -2\Delta_{\text{K}} & -3\Sigma_{\text{K}}+4D
        \end{pmatrix},
    \end{align}
    where the notation $\Delta_{\text{K}}\equiv J_{\text{K2}}-J_{\text{K3}}$ and $\Sigma_{\text{K}}\equiv J_{\text{K2}}+J_{\text{K3}} \equiv 2 J_{\text{K}}$ has been introduced. 
    When the application of the exchange coupling is isotropic by choosing $J_{\text{K2}}=J_{\text{K3}}=J_{\text{K}}$, $\Delta_{\text{K}}=0$, and the $\ket{\uparrow}\ket{1,1}$ state is no longer coupled to the other states within this block. 
    For the other two states, the effective Hamiltonian becomes, 
    \begin{align}
    \label{eqn:hamiltonian-isokondo-subblock-3d2-devicebasis}
        -\frac{1}{2}\left(D-\frac{3}{2}J_{\text{K}}\right)
        \begin{pmatrix}
            1 & 0\\
            0 & -1
        \end{pmatrix}+
        J_{\text{K}}
        \begin{pmatrix}
            0 & 1\\
            1 & 0
        \end{pmatrix}.
    \end{align}
    For comparison, the same procedure is repeated for the $m=1/2$ subspace, where the effective Hamiltonian corresponding with the $\ket{\uparrow}\ket{1,0}$ and $\ket{\downarrow}\ket{1,1}$ basis takes the form,
    \begin{align}
    \label{eqn:hamiltonian-isokondo-subblock-1d2-devicebasis}
        \frac{1}{2}\left(D+\frac{1}{2}J_{\text{K}}\right)
        \begin{pmatrix}
            1 & 0\\
            0 & -1
        \end{pmatrix}+
        \frac{1}{\sqrt{2}}J_{\text{K}}
        \begin{pmatrix}
            0 & 1\\
            1 & 0
        \end{pmatrix}.
    \end{align}
    
    Assuming that the initial density matrix of the system is fully populated in the $\ket{\downarrow}\ket{2,2}$ state, an application of the Rabi formula results in the probability of measuring the $\ket{\uparrow}\ket{2,1}$ state as,
    \begin{align}
    P_{\ket{\uparrow}\ket{2,+1}}(t) &= \left(\frac{J_{\text{K}}}{\Omega}\right)^{2}\sin^{2}(\Omega t),
    \end{align}
    with Rabi frequency,
    \begin{align}
        \Omega &= \sqrt{J_{\text{K}}^{2}+\frac{1}{4}\left(D-\frac{3}{2}J_{\text{K}}\right)^{2}}.
    \end{align}
    Transforming between the considered device basis states and their site-basis representation ($\ket{m_{1}}\ket{s_{23},m_{23}}\rightarrow\ket{m_{1}}\ket{m_{2}}\ket{m_{3}}$), 
    \begin{align}
    \label{eqn:device2site-s23-1-transition}
        \ket{\downarrow}\ket{2,2} &= \ket{\downarrow}\ket{1}\ket{1},\\
        \ket{\uparrow}\ket{2,1} &= \frac{1}{\sqrt{2}}\left( \ket{\uparrow}\ket{0}\ket{1}+\ket{\uparrow}\ket{1}\ket{0}\right),
    \end{align}
    one can see that the $\ket{\downarrow}\ket{2,2}$ state corresponds with a non-entangled coupled particle state, and the $\ket{\uparrow}\ket{2,1}$ corresponds with a maximally entangled coupled particle state. 
    Thus a single measurement of particle 1's spin orientation determines the entanglement state of particle 2 and 3. 
    This demonstrates the read out of the entanglement state if the measurement of the $S_{1}$ spin polarization is taken at any general time $t$. 
    This also demonstrates preparation of the entanglement state if the $S_{1}$ spin polarization is measured at a specific time $t$ corresponding with a peak in the Rabi oscillation.
    
    As shown in Eq.~(\ref{eqn:hamiltonian-isokondo-subblock-3d2-devicebasis}) and Eq.~(\ref{eqn:hamiltonian-isokondo-subblock-1d2-devicebasis}), both the magnetic anisotropy $D$ and average exchange interaction strength $J_{\text{K}}$ determine the Rabi frequency and transition amplitudes of the system. 
    When the Rabi frequencies and amplitudes are calculated for the other possible two state systems in the $S_{2,3}=1$ model, we see that particular conditions on the magnitude and sign of $J_{\text{K}}$ and $D$ result in resonant transition probabilities, i.e. each state's transition probability oscillates with a maximum amplitude of 1. 
    Table \ref{tab:resonance-s1} lists these possible magnetic anisotropy and exchange strength resonance conditions, i.e. DJ resonances, for two-state switching.
    \begin{table}
        \caption{\label{tab:resonance-s1}
        Pure state transitions for the $S_{2,3}=1$ model, where $J_{\text{R}}$ is the condition on $J_{\text{K}}$ to reach resonance, $P_{R}$ is the maximum transition probability amplitude at resonance, and $\Omega_{R}$ is the Rabi frequency at resonance.}
        \begin{ruledtabular}
        \begin{tabular}{llll}
        State Transitions & $J_{\text{R}}$ & $P_{\text{R}}$ & $\Omega_{\text{R}}$ \\
        
        $\ket{\uparrow}\ket{2,+1},\;\ket{\downarrow}\ket{2,+2}$ & $\frac{2}{3}D$ & $1$ & $\frac{2}{3}\abs{D}$\\
        
        $\ket{\uparrow}\ket{2,-2},\;\ket{\downarrow}\ket{2,-1}$ & $\frac{2}{3}D$ & $1$ & $\frac{2}{3}\abs{D}$\\
       
        $\ket{\uparrow}\ket{1,0},\;\ket{\downarrow}\ket{1,+1}$ & $-2D$ & $1$ & $\sqrt{2}\abs{D}$\\
       
        $\ket{\uparrow}\ket{1,-1},\;\ket{\downarrow}\ket{1,0}$ & $-2D$ & $1$ & $\sqrt{2}\abs{D}$
        \end{tabular}
        \end{ruledtabular}
    \end{table}

    To see the physical consequence of these DJ resonances, we turn to a representation of the states involved in a transition, where one can project the two-state systems onto a Bloch sphere. 
    For the $m=3/2$ case, Eq.~(\ref{eqn:hamiltonian-isokondo-subblock-3d2-devicebasis}) is written suggestively to highlight the effect of the unitary operator $U(t) = e^{-i\mathcal{H}t}$ on the Bloch vector $\vb{V}$ prepared as $(\abs{\vb{V}},\theta,\phi)=(1,0,0)$. 
    In the case of Eq.~(\ref{eqn:hamiltonian-isokondo-subblock-3d2-devicebasis}), the Bloch vector's poles are defined by the $\ket{\downarrow}\ket{2,2}$ and $\ket{\uparrow}\ket{2,1}$ states. 
    The first term in Eq.~(\ref{eqn:hamiltonian-isokondo-subblock-3d2-devicebasis}) corresponds with a rotation (up to a global phase) of the Bloch vector about the $z$-axis with a frequency $D-\frac{3}{2}J_{\text{K}}$ and the second with a rotation about the $x$-axis with a frequency of $2J_{\text{K}}$.
    \begin{figure}[h]
    \includegraphics[width=\columnwidth]{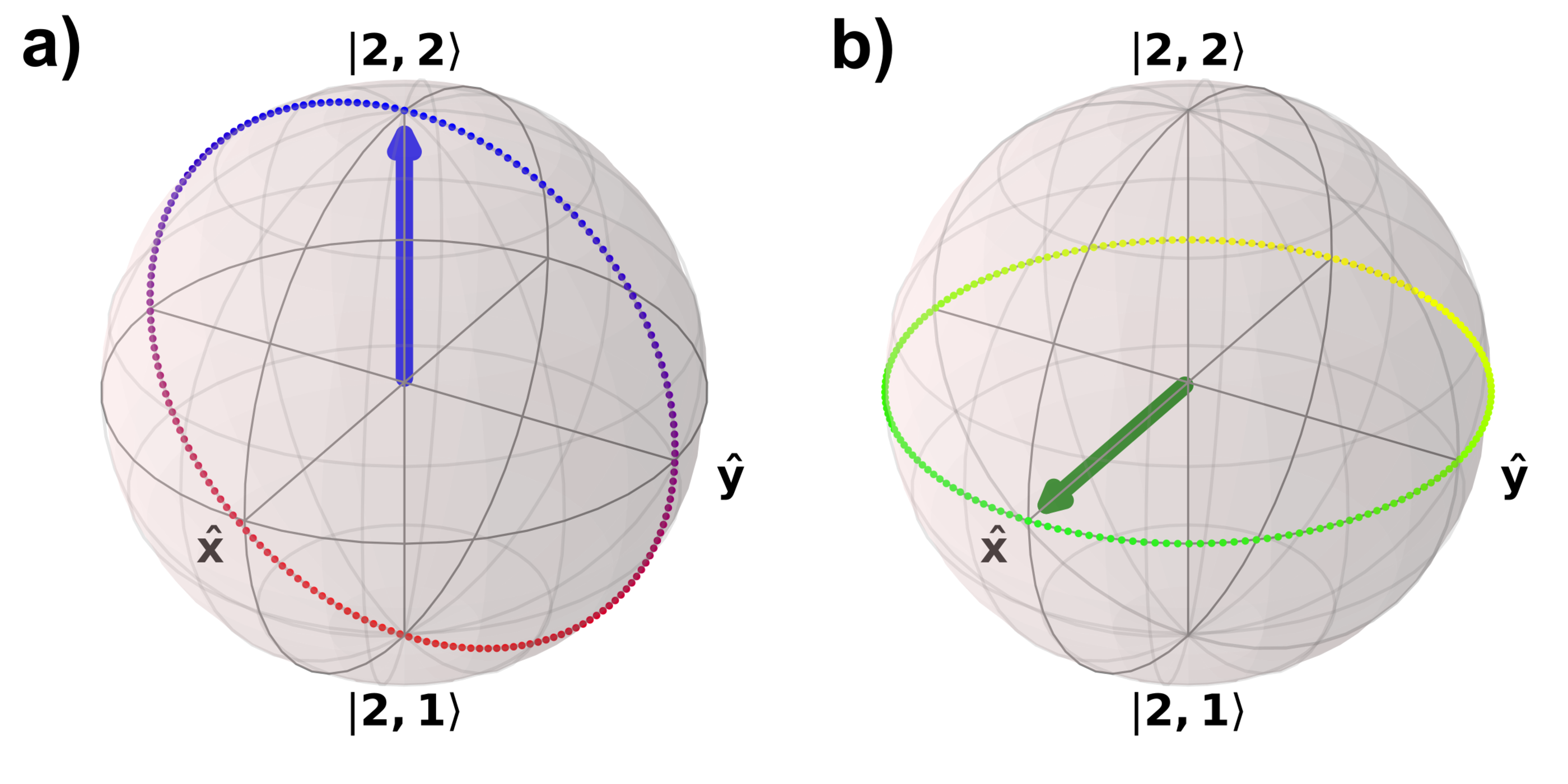}
    \colorcaption{\label{fig:bloch-zandx-rot}
    Bloch sphere representation of the states within the  $\ket{\downarrow}\ket{2,2},\ket{\uparrow}\ket{2,1}$, $m=\tfrac{3}{2}$ subspace when $J_{\text{K2}}=J_{\text{K3}}=J_{\text{K}}$. (a) When the $D$ and $J_{\text{K}}$ strengths are tuned to a DJ resonance, rotation about the x-axis is realized. (b) When $J_{\text{K}}=0$, $D$-modulated rotation about the z-axis is possible.}
    \end{figure}
    As shown in Fig~\ref{fig:bloch-zandx-rot}, at the DJ resonance condition of $J_{\text{K}}=\frac{2}{3}D$, the $z$-axis rotation vector is 0, and the Bloch vector is rotated solely about the $x$-axis. 
    In this way, the DJ resonance condition realizes control of the Bloch vector in the $x = 0$ plane. 
    Physically the magnitude of $J_{\text{K}}$ and $D$ determine the contribution of the device states that are energetically favorable for that parameter. 
    When these two parameters are equal, these state contributions are equally balanced. In other words, there is equal probability to collapse the device state upon measurement to one that favors exchange coupling or to one that favors magnetic anisotropy. 
    Similarly by turning off the exchange coupling between the two $S_{2,3}$ particles, the $x$-axis rotation is suppressed, leading to rotation solely about the $z$-axis with frequency $D$.
    
\begin{figure}
    \includegraphics[width=\columnwidth]{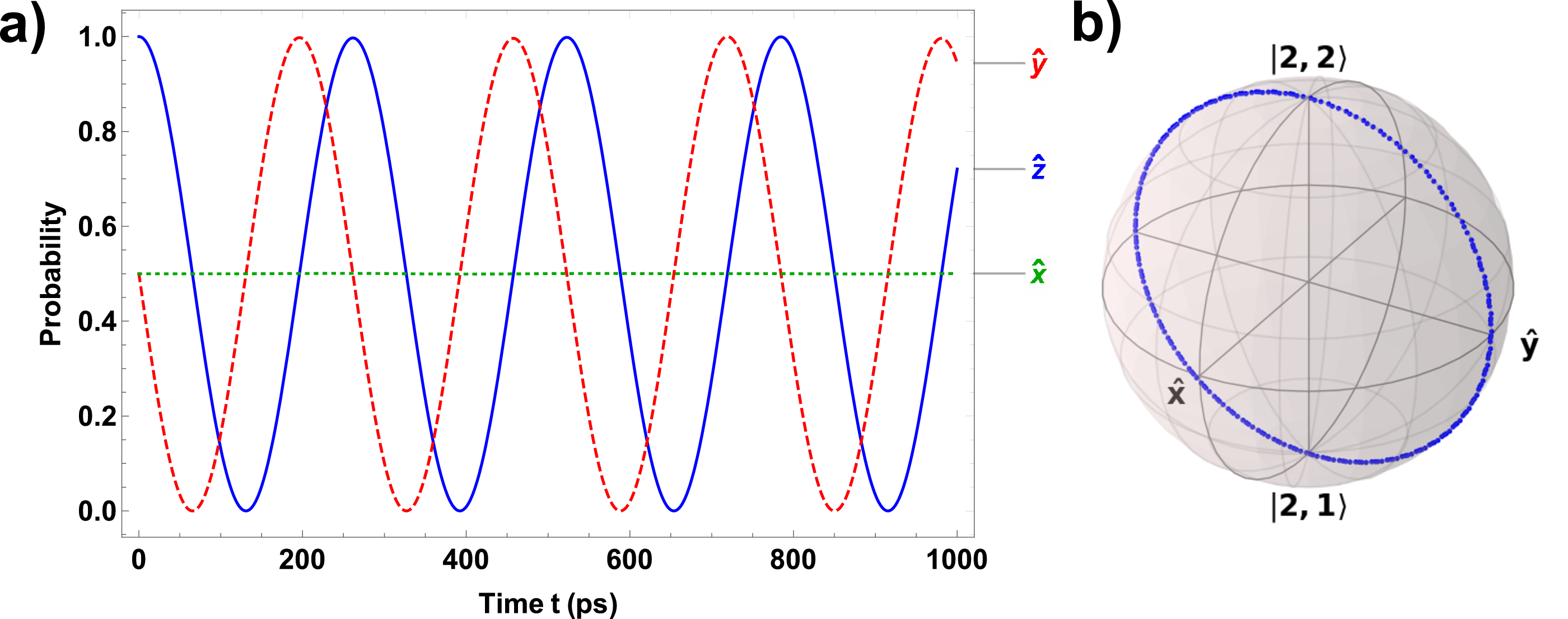}
    \colorcaption{\label{fig:spin-1-anikondo-p0995-bloch}
    (a) Probability of measuring a state corresponding with the $\hat{x}$ (dotted), $\hat{y}$ (dashed), and $\hat{z}$ (solid) unit vectors on the Bloch sphere defined in Fig.~\ref{fig:bloch-zandx-rot}, as a function of time. The Bloch vector is initially prepared in the state $\ket{\downarrow}\ket{2,2}$, and an anisotropic application of the exchange coupling strength has been used ($\Delta_{\text{K}} / J_{\text{K}}=0.072$). The parameter set has been prepared around the DJ resonance.  (b) The corresponding Bloch sphere representation of the path traced by the Bloch vector over the same time interval considered.
    In units of $\text{cm}^{-1}$, the parameters are: $J_{\text{H}}=-0.05$, $J_{\text{K}}=-0.40$, $D=-0.60$, $t=0.05$.}
    \end{figure}

    Next, we find that when small values of anisotropy of the exchange coupling are included, the numerical calculation of the Bloch vector's projection on the x-axis oscillates, resulting in a correspondingly small deviation in its z-axis projection.
    This is due to the inclusion of additional off-diagonal states (e.g. see Eq.~(\ref{eqn:hamiltonian-subblock-3d2-devicebasis})) that correspond with one of the azimuthal axes in the Bloch sphere. 
    Additionally, these additional states include contributions of the Heisenberg-like exchange Hamiltonian, so that three parameters now control the rotation of the Bloch vector. 
    Despite these contributions, we find that for certain parameters, the projection of the Bloch vector onto the z-axis (which directly corresponds with the switching behavior as measured by the electron) results in a maximal transition probability above $P=0.995$ even when using significant ratios of anisotropy in the application of the exchange coupling, in our case $\Delta_{\text{K}}/J_{\text{K}}=0.072$, as shown in Fig.~\ref{fig:spin-1-anikondo-p0995-bloch}. 
    When the anisotropic application of the exchange coupling is larger than the $\Delta_{\text{K}} / J_{\text{K}}=0.072$ ratio for the set of parameters considered, the projection onto the Bloch sphere's z-axis is more distorted, as additional rotations about axes lying in the azimuthal plane are included.
   
    The behavior of the $S_{2,3}=1$ model is not possible for $S_{2,3}=\frac{1}{2}$ without additional spin selection methods. 
    Repeating the same type of procedure and analysis for the $S_{2,3}=\frac{1}{2}$ model, and noting that magnetic anisotropy is not expected for $S=\frac{1}{2}$ particles, we find that the maximum probability amplitude is $8/9$ for both transition types.
    When the two states involved in a transition are mapped onto a Bloch sphere, the effect of the unitary operator as a rotation is not about an axis solely on the azimuthal plane, but instead contains a component in the polar plane. 
    
     Fig.~\ref{fig:spin-1-anikondo-p0995-bloch} forms the primary consequence of the main result of this paper, namely that resonance conditions exist in the $S_{2,3}=1$ model in which preparation and measurement of the coupled particles' degree of entanglement can be accomplish by appropriate measurement of the $S_{1}=\frac{1}{2}$ particle. 
     Furthermore, realization of these DJ resonances is robust against anisotropy of the exchange coupling between the coupled particles. This is found to not be the case for the $S_{2,3}=\frac{1}{2}$ model. 
     In particular, as shown in Table~\ref{tab:resonance-s1}, the DJ resonance conditions for maximal transitions between non-entangled and entangled states is controlled by the non-trivial interaction of the exchange coupling, $J_{\text{K2}}$ and $J_{\text{K3}}$, between particle 1 and 2/3 and the magnetic anisotropy $D$ of particle 2 and 3 in the $S_{2,3}=1$ paradigm. 
     As far as we are aware, these resonances have not yet been uncovered.
     
     The DJ resonance conditions indicate a possible avenue for investigating complex spin spaces and the conditions required to simplify complicated spin Hamiltonians, such as those that represent the interactions of magnetic monomers or dimers with an electron. 
     It is interesting to note that because our model has not required particular physical mechanisms for the exchange coupling and magnetic anisotropy, it is possible that outside of the condensed matter paradigm, the block diagonalization of similar $\mathbb{C}^{18}$ state systems could result in isolated SO(2) representation sub-groups.
     The DJ resonance feature introduces a different level of control in Bloch vector rotation operations. 
     The inclusion of the $S_{1}=\frac{1}{2}$ particle allows for preparation, manipulation, and reading of the entangled coupled particles. 
    
     In quantum dot QIS systems, states are often prepared with applied magnetic fields.
     Electrically-controlled methods, however, are attractive because of the relative ease of manipulating electric fields within a variety of environmental conditions.
     The results of our model predict that single-electron control of entangled particles without any use of applied magnetic fields is possible. 
     Furthermore, an important conclusion to be drawn from Fig.~\ref{fig:spin-1-anikondo-p0995-bloch} is that the Bloch vector, in time, is forgiving against misalignment, so that this scheme does not actually require impossible experimental perfection to work.

    Last, we note that if the resonances are used for QIS applications, several additional factors must be incorporated that have not been considered in this paper. 
    QIS applications usually require incorporation of the electronic degrees of freedom, e.g. because of transient charged $S_{1}$ particles. By doing so, the $S_{1}$ particle's source and drain must be included. 
    If the $S_{1}$ particle is not transient, but instead is confined on a surface or within bulk material, additional exchange coupling interactions between the particle and the confinement source may need to be accounted for.
    Regardless of the physical mechanism chosen to realize this model, the use of these DJ resonances provides an exciting avenue to uncover interesting highly-correlated spin phenomena. 
    
\begin{acknowledgments}
    We thank James Freericks, Peter Dowben, Volodymyr Turkowski, Silas Hoffman, and Dave Austin for helpful discussions. This work was supported by the Center for Molecular Magnetic Quantum Materials, an Energy Frontier Research Center funded by the U.S. Department of Energy, Office of Science, Basic Energy Sciences under Award No. DE-SC0019330. The authors declare no competing financial interests.
\end{acknowledgments}


\begin{thebibliography}{29}%
\makeatletter
\providecommand \@ifxundefined [1]{%
 \@ifx{#1\undefined}
}%
\providecommand \@ifnum [1]{%
 \ifnum #1\expandafter \@firstoftwo
 \else \expandafter \@secondoftwo
 \fi
}%
\providecommand \@ifx [1]{%
 \ifx #1\expandafter \@firstoftwo
 \else \expandafter \@secondoftwo
 \fi
}%
\providecommand \natexlab [1]{#1}%
\providecommand \enquote  [1]{``#1''}%
\providecommand \bibnamefont  [1]{#1}%
\providecommand \bibfnamefont [1]{#1}%
\providecommand \citenamefont [1]{#1}%
\providecommand \href@noop [0]{\@secondoftwo}%
\providecommand \href [0]{\begingroup \@sanitize@url \@href}%
\providecommand \@href[1]{\@@startlink{#1}\@@href}%
\providecommand \@@href[1]{\endgroup#1\@@endlink}%
\providecommand \@sanitize@url [0]{\catcode `\\12\catcode `\$12\catcode
  `\&12\catcode `\#12\catcode `\^12\catcode `\_12\catcode `\%12\relax}%
\providecommand \@@startlink[1]{}%
\providecommand \@@endlink[0]{}%
\providecommand \url  [0]{\begingroup\@sanitize@url \@url }%
\providecommand \@url [1]{\endgroup\@href {#1}{\urlprefix }}%
\providecommand \urlprefix  [0]{URL }%
\providecommand \Eprint [0]{\href }%
\providecommand \doibase [0]{https://doi.org/}%
\providecommand \selectlanguage [0]{\@gobble}%
\providecommand \bibinfo  [0]{\@secondoftwo}%
\providecommand \bibfield  [0]{\@secondoftwo}%
\providecommand \translation [1]{[#1]}%
\providecommand \BibitemOpen [0]{}%
\providecommand \bibitemStop [0]{}%
\providecommand \bibitemNoStop [0]{.\EOS\space}%
\providecommand \EOS [0]{\spacefactor3000\relax}%
\providecommand \BibitemShut  [1]{\csname bibitem#1\endcsname}%
\let\auto@bib@innerbib\@empty
\bibitem [{\citenamefont {Nielsen}\ and\ \citenamefont
  {Chuang}(2010)}]{nielsenbook}%
  \BibitemOpen
  \bibfield  {author} {\bibinfo {author} {\bibfnamefont {M.~A.}\ \bibnamefont
  {Nielsen}}\ and\ \bibinfo {author} {\bibfnamefont {I.~L.}\ \bibnamefont
  {Chuang}},\ }\href {https://doi.org/DOI: 10.1017/CBO9780511976667} {\emph
  {\bibinfo {title} {Quantum Computation and Quantum Information: 10th
  Anniversary Edition}}}\ (\bibinfo  {publisher} {Cambridge University Press},\
  \bibinfo {address} {Cambridge},\ \bibinfo {year} {2010})\BibitemShut
  {NoStop}%
\bibitem [{\citenamefont {Loss}\ and\ \citenamefont
  {DiVincenzo}(1998)}]{loss98}%
  \BibitemOpen
  \bibfield  {author} {\bibinfo {author} {\bibfnamefont {D.}~\bibnamefont
  {Loss}}\ and\ \bibinfo {author} {\bibfnamefont {D.~P.}\ \bibnamefont
  {DiVincenzo}},\ }\bibfield  {title} {\bibinfo {title} {Quantum computation
  with quantum dots},\ }\href {https://doi.org/10.1103/PhysRevA.57.120}
  {\bibfield  {journal} {\bibinfo  {journal} {Physical Review A}\ }\textbf
  {\bibinfo {volume} {57}},\ \bibinfo {pages} {120} (\bibinfo {year}
  {1998})}\BibitemShut {NoStop}%
\bibitem [{\citenamefont {Petta}\ \emph {et~al.}(2005)\citenamefont {Petta},
  \citenamefont {Johnson}, \citenamefont {Taylor}, \citenamefont {Laird},
  \citenamefont {Yacoby}, \citenamefont {Lukin}, \citenamefont {Marcus},
  \citenamefont {Hanson},\ and\ \citenamefont {Gossard}}]{petta05}%
  \BibitemOpen
  \bibfield  {author} {\bibinfo {author} {\bibfnamefont {J.~R.}\ \bibnamefont
  {Petta}}, \bibinfo {author} {\bibfnamefont {A.~C.}\ \bibnamefont {Johnson}},
  \bibinfo {author} {\bibfnamefont {J.~M.}\ \bibnamefont {Taylor}}, \bibinfo
  {author} {\bibfnamefont {E.~A.}\ \bibnamefont {Laird}}, \bibinfo {author}
  {\bibfnamefont {A.}~\bibnamefont {Yacoby}}, \bibinfo {author} {\bibfnamefont
  {M.~D.}\ \bibnamefont {Lukin}}, \bibinfo {author} {\bibfnamefont {C.~M.}\
  \bibnamefont {Marcus}}, \bibinfo {author} {\bibfnamefont {M.~P.}\
  \bibnamefont {Hanson}},\ and\ \bibinfo {author} {\bibfnamefont {A.~C.}\
  \bibnamefont {Gossard}},\ }\bibfield  {title} {\bibinfo {title} {Coherent
  manipulation of coupled electron spins in semiconductor quantum dots},\
  }\href {https://doi.org/10.1126/science.1116955} {\bibfield  {journal}
  {\bibinfo  {journal} {Science}\ }\textbf {\bibinfo {volume} {309}},\ \bibinfo
  {pages} {2180} (\bibinfo {year} {2005})}\BibitemShut {NoStop}%
\bibitem [{\citenamefont {Hanson}\ \emph {et~al.}(2007)\citenamefont {Hanson},
  \citenamefont {Kouwenhoven}, \citenamefont {Petta}, \citenamefont {Tarucha},\
  and\ \citenamefont {Vandersypen}}]{hanson07}%
  \BibitemOpen
  \bibfield  {author} {\bibinfo {author} {\bibfnamefont {R.}~\bibnamefont
  {Hanson}}, \bibinfo {author} {\bibfnamefont {L.~P.}\ \bibnamefont
  {Kouwenhoven}}, \bibinfo {author} {\bibfnamefont {J.~R.}\ \bibnamefont
  {Petta}}, \bibinfo {author} {\bibfnamefont {S.}~\bibnamefont {Tarucha}},\
  and\ \bibinfo {author} {\bibfnamefont {L.~M.~K.}\ \bibnamefont
  {Vandersypen}},\ }\bibfield  {title} {\bibinfo {title} {Spins in few-electron
  quantum dots},\ }\href {https://doi.org/10.1103/RevModPhys.79.1217}
  {\bibfield  {journal} {\bibinfo  {journal} {Reviews of Modern Physics}\
  }\textbf {\bibinfo {volume} {79}},\ \bibinfo {pages} {1217} (\bibinfo {year}
  {2007})}\BibitemShut {NoStop}%
\bibitem [{\citenamefont {Noiri}\ \emph {et~al.}(2016)\citenamefont {Noiri},
  \citenamefont {Yoneda}, \citenamefont {Nakajima}, \citenamefont {Otsuka},
  \citenamefont {Delbecq}, \citenamefont {Takeda}, \citenamefont {Amaha},
  \citenamefont {Allison}, \citenamefont {Ludwig}, \citenamefont {Wieck},\ and\
  \citenamefont {Tarucha}}]{noiri16}%
  \BibitemOpen
  \bibfield  {author} {\bibinfo {author} {\bibfnamefont {A.}~\bibnamefont
  {Noiri}}, \bibinfo {author} {\bibfnamefont {J.}~\bibnamefont {Yoneda}},
  \bibinfo {author} {\bibfnamefont {T.}~\bibnamefont {Nakajima}}, \bibinfo
  {author} {\bibfnamefont {T.}~\bibnamefont {Otsuka}}, \bibinfo {author}
  {\bibfnamefont {M.~R.}\ \bibnamefont {Delbecq}}, \bibinfo {author}
  {\bibfnamefont {K.}~\bibnamefont {Takeda}}, \bibinfo {author} {\bibfnamefont
  {S.}~\bibnamefont {Amaha}}, \bibinfo {author} {\bibfnamefont
  {G.}~\bibnamefont {Allison}}, \bibinfo {author} {\bibfnamefont
  {A.}~\bibnamefont {Ludwig}}, \bibinfo {author} {\bibfnamefont {A.~D.}\
  \bibnamefont {Wieck}},\ and\ \bibinfo {author} {\bibfnamefont
  {S.}~\bibnamefont {Tarucha}},\ }\bibfield  {title} {\bibinfo {title}
  {Coherent electron-spin-resonance manipulation of three individual spins in a
  triple quantum dot},\ }\href {https://doi.org/10.1063/1.4945592} {\bibfield
  {journal} {\bibinfo  {journal} {Applied Physics Letters}\ }\textbf {\bibinfo
  {volume} {108}},\ \bibinfo {pages} {153101} (\bibinfo {year}
  {2016})}\BibitemShut {NoStop}%
\bibitem [{\citenamefont {Noiri}\ \emph {et~al.}(2018)\citenamefont {Noiri},
  \citenamefont {Nakajima}, \citenamefont {Yoneda}, \citenamefont {Delbecq},
  \citenamefont {Stano}, \citenamefont {Otsuka}, \citenamefont {Takeda},
  \citenamefont {Amaha}, \citenamefont {Allison}, \citenamefont {Kawasaki},
  \citenamefont {Kojima}, \citenamefont {Ludwig}, \citenamefont {Wieck},
  \citenamefont {Loss},\ and\ \citenamefont {Tarucha}}]{noiri18}%
  \BibitemOpen
  \bibfield  {author} {\bibinfo {author} {\bibfnamefont {A.}~\bibnamefont
  {Noiri}}, \bibinfo {author} {\bibfnamefont {T.}~\bibnamefont {Nakajima}},
  \bibinfo {author} {\bibfnamefont {J.}~\bibnamefont {Yoneda}}, \bibinfo
  {author} {\bibfnamefont {M.~R.}\ \bibnamefont {Delbecq}}, \bibinfo {author}
  {\bibfnamefont {P.}~\bibnamefont {Stano}}, \bibinfo {author} {\bibfnamefont
  {T.}~\bibnamefont {Otsuka}}, \bibinfo {author} {\bibfnamefont
  {K.}~\bibnamefont {Takeda}}, \bibinfo {author} {\bibfnamefont
  {S.}~\bibnamefont {Amaha}}, \bibinfo {author} {\bibfnamefont
  {G.}~\bibnamefont {Allison}}, \bibinfo {author} {\bibfnamefont
  {K.}~\bibnamefont {Kawasaki}}, \bibinfo {author} {\bibfnamefont
  {Y.}~\bibnamefont {Kojima}}, \bibinfo {author} {\bibfnamefont
  {A.}~\bibnamefont {Ludwig}}, \bibinfo {author} {\bibfnamefont {A.~D.}\
  \bibnamefont {Wieck}}, \bibinfo {author} {\bibfnamefont {D.}~\bibnamefont
  {Loss}},\ and\ \bibinfo {author} {\bibfnamefont {S.}~\bibnamefont
  {Tarucha}},\ }\bibfield  {title} {\bibinfo {title} {A fast quantum interface
  between different spin qubit encodings},\ }\href
  {https://doi.org/10.1038/s41467-018-07522-1} {\bibfield  {journal} {\bibinfo
  {journal} {Nature Communications}\ }\textbf {\bibinfo {volume} {9}},\
  \bibinfo {pages} {5066} (\bibinfo {year} {2018})}\BibitemShut {NoStop}%
\bibitem [{\citenamefont {Nakajima}\ \emph {et~al.}(2019)\citenamefont
  {Nakajima}, \citenamefont {Noiri}, \citenamefont {Yoneda}, \citenamefont
  {Delbecq}, \citenamefont {Stano}, \citenamefont {Otsuka}, \citenamefont
  {Takeda}, \citenamefont {Amaha}, \citenamefont {Allison}, \citenamefont
  {Kawasaki}, \citenamefont {Ludwig}, \citenamefont {Wieck}, \citenamefont
  {Loss},\ and\ \citenamefont {Tarucha}}]{nakajima19}%
  \BibitemOpen
  \bibfield  {author} {\bibinfo {author} {\bibfnamefont {T.}~\bibnamefont
  {Nakajima}}, \bibinfo {author} {\bibfnamefont {A.}~\bibnamefont {Noiri}},
  \bibinfo {author} {\bibfnamefont {J.}~\bibnamefont {Yoneda}}, \bibinfo
  {author} {\bibfnamefont {M.~R.}\ \bibnamefont {Delbecq}}, \bibinfo {author}
  {\bibfnamefont {P.}~\bibnamefont {Stano}}, \bibinfo {author} {\bibfnamefont
  {T.}~\bibnamefont {Otsuka}}, \bibinfo {author} {\bibfnamefont
  {K.}~\bibnamefont {Takeda}}, \bibinfo {author} {\bibfnamefont
  {S.}~\bibnamefont {Amaha}}, \bibinfo {author} {\bibfnamefont
  {G.}~\bibnamefont {Allison}}, \bibinfo {author} {\bibfnamefont
  {K.}~\bibnamefont {Kawasaki}}, \bibinfo {author} {\bibfnamefont
  {A.}~\bibnamefont {Ludwig}}, \bibinfo {author} {\bibfnamefont {A.~D.}\
  \bibnamefont {Wieck}}, \bibinfo {author} {\bibfnamefont {D.}~\bibnamefont
  {Loss}},\ and\ \bibinfo {author} {\bibfnamefont {S.}~\bibnamefont
  {Tarucha}},\ }\bibfield  {title} {\bibinfo {title} {Quantum non-demolition
  measurement of an electron spin qubit},\ }\href
  {https://doi.org/10.1038/s41565-019-0426-x} {\bibfield  {journal} {\bibinfo
  {journal} {Nature Nanotechnology}\ }\textbf {\bibinfo {volume} {14}},\
  \bibinfo {pages} {555} (\bibinfo {year} {2019})}\BibitemShut {NoStop}%
\bibitem [{\citenamefont {Yang}\ \emph {et~al.}(2020)\citenamefont {Yang},
  \citenamefont {Leon}, \citenamefont {Hwang}, \citenamefont {Saraiva},
  \citenamefont {Tanttu}, \citenamefont {Huang}, \citenamefont
  {Camirand~Lemyre}, \citenamefont {Chan}, \citenamefont {Tan}, \citenamefont
  {Hudson}, \citenamefont {Itoh}, \citenamefont {Morello}, \citenamefont
  {Pioro-Ladrière}, \citenamefont {Laucht},\ and\ \citenamefont
  {Dzurak}}]{yang2020}%
  \BibitemOpen
  \bibfield  {author} {\bibinfo {author} {\bibfnamefont {C.~H.}\ \bibnamefont
  {Yang}}, \bibinfo {author} {\bibfnamefont {R.~C.~C.}\ \bibnamefont {Leon}},
  \bibinfo {author} {\bibfnamefont {J.~C.~C.}\ \bibnamefont {Hwang}}, \bibinfo
  {author} {\bibfnamefont {A.}~\bibnamefont {Saraiva}}, \bibinfo {author}
  {\bibfnamefont {T.}~\bibnamefont {Tanttu}}, \bibinfo {author} {\bibfnamefont
  {W.}~\bibnamefont {Huang}}, \bibinfo {author} {\bibfnamefont
  {J.}~\bibnamefont {Camirand~Lemyre}}, \bibinfo {author} {\bibfnamefont
  {K.~W.}\ \bibnamefont {Chan}}, \bibinfo {author} {\bibfnamefont {K.~Y.}\
  \bibnamefont {Tan}}, \bibinfo {author} {\bibfnamefont {F.~E.}\ \bibnamefont
  {Hudson}}, \bibinfo {author} {\bibfnamefont {K.~M.}\ \bibnamefont {Itoh}},
  \bibinfo {author} {\bibfnamefont {A.}~\bibnamefont {Morello}}, \bibinfo
  {author} {\bibfnamefont {M.}~\bibnamefont {Pioro-Ladrière}}, \bibinfo
  {author} {\bibfnamefont {A.}~\bibnamefont {Laucht}},\ and\ \bibinfo {author}
  {\bibfnamefont {A.~S.}\ \bibnamefont {Dzurak}},\ }\bibfield  {title}
  {\bibinfo {title} {Operation of a silicon quantum processor unit cell above
  one kelvin},\ }\href {https://doi.org/10.1038/s41586-020-2171-6} {\bibfield
  {journal} {\bibinfo  {journal} {Nature}\ }\textbf {\bibinfo {volume} {580}},\
  \bibinfo {pages} {350} (\bibinfo {year} {2020})}\BibitemShut {NoStop}%
\bibitem [{\citenamefont {Leon}\ \emph {et~al.}(2020)\citenamefont {Leon},
  \citenamefont {Yang}, \citenamefont {Hwang}, \citenamefont {Lemyre},
  \citenamefont {Tanttu}, \citenamefont {Huang}, \citenamefont {Chan},
  \citenamefont {Tan}, \citenamefont {Hudson}, \citenamefont {Itoh},
  \citenamefont {Morello}, \citenamefont {Laucht}, \citenamefont
  {Pioro-Ladrière}, \citenamefont {Saraiva},\ and\ \citenamefont
  {Dzurak}}]{leon2020}%
  \BibitemOpen
  \bibfield  {author} {\bibinfo {author} {\bibfnamefont {R.~C.~C.}\
  \bibnamefont {Leon}}, \bibinfo {author} {\bibfnamefont {C.~H.}\ \bibnamefont
  {Yang}}, \bibinfo {author} {\bibfnamefont {J.~C.~C.}\ \bibnamefont {Hwang}},
  \bibinfo {author} {\bibfnamefont {J.~C.}\ \bibnamefont {Lemyre}}, \bibinfo
  {author} {\bibfnamefont {T.}~\bibnamefont {Tanttu}}, \bibinfo {author}
  {\bibfnamefont {W.}~\bibnamefont {Huang}}, \bibinfo {author} {\bibfnamefont
  {K.~W.}\ \bibnamefont {Chan}}, \bibinfo {author} {\bibfnamefont {K.~Y.}\
  \bibnamefont {Tan}}, \bibinfo {author} {\bibfnamefont {F.~E.}\ \bibnamefont
  {Hudson}}, \bibinfo {author} {\bibfnamefont {K.~M.}\ \bibnamefont {Itoh}},
  \bibinfo {author} {\bibfnamefont {A.}~\bibnamefont {Morello}}, \bibinfo
  {author} {\bibfnamefont {A.}~\bibnamefont {Laucht}}, \bibinfo {author}
  {\bibfnamefont {M.}~\bibnamefont {Pioro-Ladrière}}, \bibinfo {author}
  {\bibfnamefont {A.}~\bibnamefont {Saraiva}},\ and\ \bibinfo {author}
  {\bibfnamefont {A.~S.}\ \bibnamefont {Dzurak}},\ }\bibfield  {title}
  {\bibinfo {title} {Coherent spin control of s-, p-, d- and f-electrons in a
  silicon quantum dot},\ }\href {https://doi.org/10.1038/s41467-019-14053-w}
  {\bibfield  {journal} {\bibinfo  {journal} {Nature Communications}\ }\textbf
  {\bibinfo {volume} {11}},\ \bibinfo {pages} {797} (\bibinfo {year}
  {2020})}\BibitemShut {NoStop}%
\bibitem [{\citenamefont {Leuenberger}\ and\ \citenamefont
  {Loss}(2001)}]{leuenberger01}%
  \BibitemOpen
  \bibfield  {author} {\bibinfo {author} {\bibfnamefont {M.~N.}\ \bibnamefont
  {Leuenberger}}\ and\ \bibinfo {author} {\bibfnamefont {D.}~\bibnamefont
  {Loss}},\ }\bibfield  {title} {\bibinfo {title} {Quantum computing in
  molecular magnets},\ }\href {https://doi.org/10.1038/35071024} {\bibfield
  {journal} {\bibinfo  {journal} {Nature}\ }\textbf {\bibinfo {volume} {410}},\
  \bibinfo {pages} {789} (\bibinfo {year} {2001})}\BibitemShut {NoStop}%
\bibitem [{\citenamefont {Vandersypen}\ \emph {et~al.}(2001)\citenamefont
  {Vandersypen}, \citenamefont {Steffen}, \citenamefont {Breyta}, \citenamefont
  {Yannoni}, \citenamefont {Sherwood},\ and\ \citenamefont
  {Chuang}}]{vandersypen01}%
  \BibitemOpen
  \bibfield  {author} {\bibinfo {author} {\bibfnamefont {L.~M.~K.}\
  \bibnamefont {Vandersypen}}, \bibinfo {author} {\bibfnamefont
  {M.}~\bibnamefont {Steffen}}, \bibinfo {author} {\bibfnamefont
  {G.}~\bibnamefont {Breyta}}, \bibinfo {author} {\bibfnamefont {C.~S.}\
  \bibnamefont {Yannoni}}, \bibinfo {author} {\bibfnamefont {M.~H.}\
  \bibnamefont {Sherwood}},\ and\ \bibinfo {author} {\bibfnamefont {I.~L.}\
  \bibnamefont {Chuang}},\ }\bibfield  {title} {\bibinfo {title} {Experimental
  realization of shor's quantum factoring algorithm using nuclear magnetic
  resonance},\ }\href {https://doi.org/10.1038/414883a} {\bibfield  {journal}
  {\bibinfo  {journal} {Nature}\ }\textbf {\bibinfo {volume} {414}},\ \bibinfo
  {pages} {883} (\bibinfo {year} {2001})}\BibitemShut {NoStop}%
\bibitem [{\citenamefont {Vincent}\ \emph {et~al.}(2012)\citenamefont
  {Vincent}, \citenamefont {Klyatskaya}, \citenamefont {Ruben}, \citenamefont
  {Wernsdorfer},\ and\ \citenamefont {Balestro}}]{vincent12}%
  \BibitemOpen
  \bibfield  {author} {\bibinfo {author} {\bibfnamefont {R.}~\bibnamefont
  {Vincent}}, \bibinfo {author} {\bibfnamefont {S.}~\bibnamefont {Klyatskaya}},
  \bibinfo {author} {\bibfnamefont {M.}~\bibnamefont {Ruben}}, \bibinfo
  {author} {\bibfnamefont {W.}~\bibnamefont {Wernsdorfer}},\ and\ \bibinfo
  {author} {\bibfnamefont {F.}~\bibnamefont {Balestro}},\ }\bibfield  {title}
  {\bibinfo {title} {Electronic read-out of a single nuclear spin using a
  molecular spin transistor},\ }\href {https://doi.org/10.1038/nature11341}
  {\bibfield  {journal} {\bibinfo  {journal} {Nature}\ }\textbf {\bibinfo
  {volume} {488}},\ \bibinfo {pages} {357} (\bibinfo {year}
  {2012})}\BibitemShut {NoStop}%
\bibitem [{\citenamefont {Ganzhorn}\ \emph {et~al.}(2013)\citenamefont
  {Ganzhorn}, \citenamefont {Klyatskaya}, \citenamefont {Ruben},\ and\
  \citenamefont {Wernsdorfer}}]{ganzhorn13}%
  \BibitemOpen
  \bibfield  {author} {\bibinfo {author} {\bibfnamefont {M.}~\bibnamefont
  {Ganzhorn}}, \bibinfo {author} {\bibfnamefont {S.}~\bibnamefont
  {Klyatskaya}}, \bibinfo {author} {\bibfnamefont {M.}~\bibnamefont {Ruben}},\
  and\ \bibinfo {author} {\bibfnamefont {W.}~\bibnamefont {Wernsdorfer}},\
  }\bibfield  {title} {\bibinfo {title} {Strong spin–phonon coupling between
  a single-molecule magnet and a carbon nanotube nanoelectromechanical
  system},\ }\href {https://doi.org/10.1038/nnano.2012.258} {\bibfield
  {journal} {\bibinfo  {journal} {Nature Nanotechnology}\ }\textbf {\bibinfo
  {volume} {8}},\ \bibinfo {pages} {165} (\bibinfo {year} {2013})}\BibitemShut
  {NoStop}%
\bibitem [{\citenamefont {Urdampilleta}\ \emph {et~al.}(2013)\citenamefont
  {Urdampilleta}, \citenamefont {Klyatskaya}, \citenamefont {Ruben},\ and\
  \citenamefont {Wernsdorfer}}]{urdampilleta13}%
  \BibitemOpen
  \bibfield  {author} {\bibinfo {author} {\bibfnamefont {M.}~\bibnamefont
  {Urdampilleta}}, \bibinfo {author} {\bibfnamefont {S.}~\bibnamefont
  {Klyatskaya}}, \bibinfo {author} {\bibfnamefont {M.}~\bibnamefont {Ruben}},\
  and\ \bibinfo {author} {\bibfnamefont {W.}~\bibnamefont {Wernsdorfer}},\
  }\bibfield  {title} {\bibinfo {title} {Landau-zener tunneling of a single
  tb${}^{3+}$ magnetic moment allowing the electronic read-out of a nuclear
  spin},\ }\href {https://doi.org/10.1103/PhysRevB.87.195412} {\bibfield
  {journal} {\bibinfo  {journal} {Physical Review B}\ }\textbf {\bibinfo
  {volume} {87}},\ \bibinfo {pages} {195412} (\bibinfo {year}
  {2013})}\BibitemShut {NoStop}%
\bibitem [{\citenamefont {Thiele}\ \emph {et~al.}(2014)\citenamefont {Thiele},
  \citenamefont {Balestro}, \citenamefont {Ballou}, \citenamefont {Klyatskaya},
  \citenamefont {Ruben},\ and\ \citenamefont {Wernsdorfer}}]{thiele14}%
  \BibitemOpen
  \bibfield  {author} {\bibinfo {author} {\bibfnamefont {S.}~\bibnamefont
  {Thiele}}, \bibinfo {author} {\bibfnamefont {F.}~\bibnamefont {Balestro}},
  \bibinfo {author} {\bibfnamefont {R.}~\bibnamefont {Ballou}}, \bibinfo
  {author} {\bibfnamefont {S.}~\bibnamefont {Klyatskaya}}, \bibinfo {author}
  {\bibfnamefont {M.}~\bibnamefont {Ruben}},\ and\ \bibinfo {author}
  {\bibfnamefont {W.}~\bibnamefont {Wernsdorfer}},\ }\bibfield  {title}
  {\bibinfo {title} {Electrically driven nuclear spin resonance in
  single-molecule magnets},\ }\href {https://doi.org/10.1126/science.1249802}
  {\bibfield  {journal} {\bibinfo  {journal} {Science}\ }\textbf {\bibinfo
  {volume} {344}},\ \bibinfo {pages} {1135} (\bibinfo {year}
  {2014})}\BibitemShut {NoStop}%
\bibitem [{\citenamefont {Pedersen}\ \emph {et~al.}(2016)\citenamefont
  {Pedersen}, \citenamefont {Ariciu}, \citenamefont {McAdams}, \citenamefont
  {Weihe}, \citenamefont {Bendix}, \citenamefont {Tuna},\ and\ \citenamefont
  {Piligkos}}]{pedersen16}%
  \BibitemOpen
  \bibfield  {author} {\bibinfo {author} {\bibfnamefont {K.~S.}\ \bibnamefont
  {Pedersen}}, \bibinfo {author} {\bibfnamefont {A.-M.}\ \bibnamefont
  {Ariciu}}, \bibinfo {author} {\bibfnamefont {S.}~\bibnamefont {McAdams}},
  \bibinfo {author} {\bibfnamefont {H.}~\bibnamefont {Weihe}}, \bibinfo
  {author} {\bibfnamefont {J.}~\bibnamefont {Bendix}}, \bibinfo {author}
  {\bibfnamefont {F.}~\bibnamefont {Tuna}},\ and\ \bibinfo {author}
  {\bibfnamefont {S.}~\bibnamefont {Piligkos}},\ }\bibfield  {title} {\bibinfo
  {title} {Toward molecular 4f single-ion magnet qubits},\ }\href
  {https://doi.org/10.1021/jacs.6b02702} {\bibfield  {journal} {\bibinfo
  {journal} {Journal of the American Chemical Society}\ }\textbf {\bibinfo
  {volume} {138}},\ \bibinfo {pages} {5801} (\bibinfo {year}
  {2016})}\BibitemShut {NoStop}%
\bibitem [{\citenamefont {Najafi}\ \emph {et~al.}(2019)\citenamefont {Najafi},
  \citenamefont {Wysocki}, \citenamefont {Park}, \citenamefont {Economou},\
  and\ \citenamefont {Barnes}}]{najafi19}%
  \BibitemOpen
  \bibfield  {author} {\bibinfo {author} {\bibfnamefont {K.}~\bibnamefont
  {Najafi}}, \bibinfo {author} {\bibfnamefont {A.~L.}\ \bibnamefont {Wysocki}},
  \bibinfo {author} {\bibfnamefont {K.}~\bibnamefont {Park}}, \bibinfo {author}
  {\bibfnamefont {S.~E.}\ \bibnamefont {Economou}},\ and\ \bibinfo {author}
  {\bibfnamefont {E.}~\bibnamefont {Barnes}},\ }\bibfield  {title} {\bibinfo
  {title} {Toward long-range entanglement between electrically driven
  single-molecule magnets},\ }\href
  {https://doi.org/10.1021/acs.jpclett.9b03131} {\bibfield  {journal} {\bibinfo
   {journal} {The Journal of Physical Chemistry Letters}\ }\textbf {\bibinfo
  {volume} {10}},\ \bibinfo {pages} {7347} (\bibinfo {year}
  {2019})}\BibitemShut {NoStop}%
\bibitem [{\citenamefont {Bogani}\ and\ \citenamefont
  {Wernsdorfer}(2008)}]{bogani08}%
  \BibitemOpen
  \bibfield  {author} {\bibinfo {author} {\bibfnamefont {L.}~\bibnamefont
  {Bogani}}\ and\ \bibinfo {author} {\bibfnamefont {W.}~\bibnamefont
  {Wernsdorfer}},\ }\bibfield  {title} {\bibinfo {title} {Molecular spintronics
  using single-molecule magnets},\ }\href {https://doi.org/10.1038/nmat2133}
  {\bibfield  {journal} {\bibinfo  {journal} {Nature Materials}\ }\textbf
  {\bibinfo {volume} {7}},\ \bibinfo {pages} {179} (\bibinfo {year}
  {2008})}\BibitemShut {NoStop}%
\bibitem [{\citenamefont {Cronenwett}\ \emph {et~al.}(1998)\citenamefont
  {Cronenwett}, \citenamefont {Oosterkamp},\ and\ \citenamefont
  {Kouwenhoven}}]{cronenwett98}%
  \BibitemOpen
  \bibfield  {author} {\bibinfo {author} {\bibfnamefont {S.~M.}\ \bibnamefont
  {Cronenwett}}, \bibinfo {author} {\bibfnamefont {T.~H.}\ \bibnamefont
  {Oosterkamp}},\ and\ \bibinfo {author} {\bibfnamefont {L.~P.}\ \bibnamefont
  {Kouwenhoven}},\ }\bibfield  {title} {\bibinfo {title} {A tunable kondo
  effect in quantum dots},\ }\href
  {https://doi.org/10.1126/science.281.5376.540} {\bibfield  {journal}
  {\bibinfo  {journal} {Science}\ }\textbf {\bibinfo {volume} {281}},\ \bibinfo
  {pages} {540} (\bibinfo {year} {1998})}\BibitemShut {NoStop}%
\bibitem [{\citenamefont {Leuenberger}\ and\ \citenamefont
  {Mucciolo}(2006)}]{leuenberger06}%
  \BibitemOpen
  \bibfield  {author} {\bibinfo {author} {\bibfnamefont {M.~N.}\ \bibnamefont
  {Leuenberger}}\ and\ \bibinfo {author} {\bibfnamefont {E.~R.}\ \bibnamefont
  {Mucciolo}},\ }\bibfield  {title} {\bibinfo {title} {Berry-phase oscillations
  of the kondo effect in single-molecule magnets},\ }\href
  {https://doi.org/10.1103/PhysRevLett.97.126601} {\bibfield  {journal}
  {\bibinfo  {journal} {Physical Review Letters}\ }\textbf {\bibinfo {volume}
  {97}},\ \bibinfo {pages} {126601} (\bibinfo {year} {2006})}\BibitemShut
  {NoStop}%
\bibitem [{\citenamefont {Gonz\'{a}lez}\ \emph {et~al.}(2008)\citenamefont
  {Gonz\'{a}lez}, \citenamefont {Leuenberger},\ and\ \citenamefont
  {Mucciolo}}]{gonzalez08}%
  \BibitemOpen
  \bibfield  {author} {\bibinfo {author} {\bibfnamefont {G.}~\bibnamefont
  {Gonz\'{a}lez}}, \bibinfo {author} {\bibfnamefont {M.~N.}\ \bibnamefont
  {Leuenberger}},\ and\ \bibinfo {author} {\bibfnamefont {E.~R.}\ \bibnamefont
  {Mucciolo}},\ }\bibfield  {title} {\bibinfo {title} {Kondo effect in
  single-molecule magnet transistors},\ }\href
  {https://doi.org/10.1103/PhysRevB.78.054445} {\bibfield  {journal} {\bibinfo
  {journal} {Physical Review B}\ }\textbf {\bibinfo {volume} {78}},\ \bibinfo
  {pages} {054445} (\bibinfo {year} {2008})}\BibitemShut {NoStop}%
\bibitem [{\citenamefont {Kondo}(1964)}]{kondo64}%
  \BibitemOpen
  \bibfield  {author} {\bibinfo {author} {\bibfnamefont {J.}~\bibnamefont
  {Kondo}},\ }\bibfield  {title} {\bibinfo {title} {Resistance minimum in
  dilute magnetic alloys},\ }\href {https://doi.org/10.1143/ptp.32.37}
  {\bibfield  {journal} {\bibinfo  {journal} {Progress of Theoretical Physics}\
  }\textbf {\bibinfo {volume} {32}},\ \bibinfo {pages} {37} (\bibinfo {year}
  {1964})}\BibitemShut {NoStop}%
\bibitem [{\citenamefont {Anderson}(1961)}]{anderson61}%
  \BibitemOpen
  \bibfield  {author} {\bibinfo {author} {\bibfnamefont {P.~W.}\ \bibnamefont
  {Anderson}},\ }\bibfield  {title} {\bibinfo {title} {Localized magnetic
  states in metals},\ }\href {https://doi.org/10.1103/PhysRev.124.41}
  {\bibfield  {journal} {\bibinfo  {journal} {Physical Review}\ }\textbf
  {\bibinfo {volume} {124}},\ \bibinfo {pages} {41} (\bibinfo {year}
  {1961})}\BibitemShut {NoStop}%
\bibitem [{\citenamefont {\v{Z}itko}\ \emph {et~al.}(2008)\citenamefont
  {\v{Z}itko}, \citenamefont {Peters},\ and\ \citenamefont
  {Pruschke}}]{zitko08}%
  \BibitemOpen
  \bibfield  {author} {\bibinfo {author} {\bibfnamefont {R.}~\bibnamefont
  {\v{Z}itko}}, \bibinfo {author} {\bibfnamefont {R.}~\bibnamefont {Peters}},\
  and\ \bibinfo {author} {\bibfnamefont {T.}~\bibnamefont {Pruschke}},\
  }\bibfield  {title} {\bibinfo {title} {Properties of anisotropic magnetic
  impurities on surfaces},\ }\href {https://doi.org/10.1103/PhysRevB.78.224404}
  {\bibfield  {journal} {\bibinfo  {journal} {Physical Review B}\ }\textbf
  {\bibinfo {volume} {78}},\ \bibinfo {pages} {224404} (\bibinfo {year}
  {2008})}\BibitemShut {NoStop}%
\bibitem [{\citenamefont {Mehl}\ and\ \citenamefont
  {DiVincenzo}(2015)}]{mehl15}%
  \BibitemOpen
  \bibfield  {author} {\bibinfo {author} {\bibfnamefont {S.}~\bibnamefont
  {Mehl}}\ and\ \bibinfo {author} {\bibfnamefont {D.~P.}\ \bibnamefont
  {DiVincenzo}},\ }\bibfield  {title} {\bibinfo {title} {Simple operation
  sequences to couple and interchange quantum information between spin qubits
  of different kinds},\ }\href {https://doi.org/10.1103/PhysRevB.92.115448}
  {\bibfield  {journal} {\bibinfo  {journal} {Physical Review B}\ }\textbf
  {\bibinfo {volume} {92}},\ \bibinfo {pages} {115448} (\bibinfo {year}
  {2015})}\BibitemShut {NoStop}%
\bibitem [{\citenamefont {Costa}\ \emph {et~al.}(2006)\citenamefont {Costa},
  \citenamefont {Bose},\ and\ \citenamefont {Omar}}]{costa06}%
  \BibitemOpen
  \bibfield  {author} {\bibinfo {author} {\bibfnamefont {A.~T.}\ \bibnamefont
  {Costa}}, \bibinfo {author} {\bibfnamefont {S.}~\bibnamefont {Bose}},\ and\
  \bibinfo {author} {\bibfnamefont {Y.}~\bibnamefont {Omar}},\ }\bibfield
  {title} {\bibinfo {title} {Entanglement of two impurities through electron
  scattering},\ }\href {https://doi.org/10.1103/PhysRevLett.96.230501}
  {\bibfield  {journal} {\bibinfo  {journal} {Physical Review Letters}\
  }\textbf {\bibinfo {volume} {96}},\ \bibinfo {pages} {230501} (\bibinfo
  {year} {2006})}\BibitemShut {NoStop}%
\bibitem [{\citenamefont {Ciccarello}\ \emph {et~al.}(2006)\citenamefont
  {Ciccarello}, \citenamefont {Palma}, \citenamefont {Zarcone}, \citenamefont
  {Omar},\ and\ \citenamefont {Vieira}}]{ciccarello06}%
  \BibitemOpen
  \bibfield  {author} {\bibinfo {author} {\bibfnamefont {F.}~\bibnamefont
  {Ciccarello}}, \bibinfo {author} {\bibfnamefont {G.~M.}\ \bibnamefont
  {Palma}}, \bibinfo {author} {\bibfnamefont {M.}~\bibnamefont {Zarcone}},
  \bibinfo {author} {\bibfnamefont {Y.}~\bibnamefont {Omar}},\ and\ \bibinfo
  {author} {\bibfnamefont {V.~R.}\ \bibnamefont {Vieira}},\ }\bibfield  {title}
  {\bibinfo {title} {Entanglement controlled single-electron transmittivity},\
  }\href {https://doi.org/10.1088/1367-2630/8/9/214} {\bibfield  {journal}
  {\bibinfo  {journal} {New Journal of Physics}\ }\textbf {\bibinfo {volume}
  {8}},\ \bibinfo {pages} {214} (\bibinfo {year} {2006})}\BibitemShut {NoStop}%
\bibitem [{\citenamefont {Ciccarello}\ \emph {et~al.}(2009)\citenamefont
  {Ciccarello}, \citenamefont {Massimo~Palma}, \citenamefont {Paternostro},
  \citenamefont {Zarcone},\ and\ \citenamefont {Omar}}]{ciccarello09}%
  \BibitemOpen
  \bibfield  {author} {\bibinfo {author} {\bibfnamefont {F.}~\bibnamefont
  {Ciccarello}}, \bibinfo {author} {\bibfnamefont {G.}~\bibnamefont
  {Massimo~Palma}}, \bibinfo {author} {\bibfnamefont {M.}~\bibnamefont
  {Paternostro}}, \bibinfo {author} {\bibfnamefont {M.}~\bibnamefont
  {Zarcone}},\ and\ \bibinfo {author} {\bibfnamefont {Y.}~\bibnamefont
  {Omar}},\ }\bibfield  {title} {\bibinfo {title} {Entanglement generation
  between two spin-s magnetic impurities in a solid via electron scattering},\
  }\href
  {https://doi.org/https://doi.org/10.1016/j.solidstatesciences.2008.01.012}
  {\bibfield  {journal} {\bibinfo  {journal} {Solid State Sciences}\ }\textbf
  {\bibinfo {volume} {11}},\ \bibinfo {pages} {931} (\bibinfo {year}
  {2009})}\BibitemShut {NoStop}%
\bibitem [{\citenamefont {Hiraoka}\ \emph {et~al.}(2017)\citenamefont
  {Hiraoka}, \citenamefont {Minamitani}, \citenamefont {Arafune}, \citenamefont
  {Tsukahara}, \citenamefont {Watanabe}, \citenamefont {Kawai},\ and\
  \citenamefont {Takagi}}]{hiraoka17}%
  \BibitemOpen
  \bibfield  {author} {\bibinfo {author} {\bibfnamefont {R.}~\bibnamefont
  {Hiraoka}}, \bibinfo {author} {\bibfnamefont {E.}~\bibnamefont {Minamitani}},
  \bibinfo {author} {\bibfnamefont {R.}~\bibnamefont {Arafune}}, \bibinfo
  {author} {\bibfnamefont {N.}~\bibnamefont {Tsukahara}}, \bibinfo {author}
  {\bibfnamefont {S.}~\bibnamefont {Watanabe}}, \bibinfo {author}
  {\bibfnamefont {M.}~\bibnamefont {Kawai}},\ and\ \bibinfo {author}
  {\bibfnamefont {N.}~\bibnamefont {Takagi}},\ }\bibfield  {title} {\bibinfo
  {title} {Single-molecule quantum dot as a kondo simulator},\ }\href
  {https://doi.org/10.1038/ncomms16012} {\bibfield  {journal} {\bibinfo
  {journal} {Nature Communications}\ }\textbf {\bibinfo {volume} {8}},\
  \bibinfo {pages} {16012} (\bibinfo {year} {2017})}\BibitemShut {NoStop}%
\end{thebibliography}
%
\end{document}